\documentclass[journal]{IEEEtai}

\usepackage[colorlinks,urlcolor=blue,linkcolor=blue,citecolor=blue]{hyperref}

\usepackage{color,array}

\usepackage{graphicx}

\IEEEoverridecommandlockouts                              

\usepackage{cite}
\usepackage{amsmath,amssymb,amsfonts}
\usepackage{algorithmicx}
\usepackage{graphicx}
\usepackage{textcomp}
\usepackage{times}
\usepackage{indentfirst}
\usepackage{latexsym}
\usepackage{algorithm}
\usepackage{algpseudocode}
\usepackage{amsmath}
\usepackage{physics}
\usepackage{amsfonts}
\usepackage{hyperref}
\hypersetup{
    colorlinks=true,
    linkcolor=blue,
    filecolor=magenta,      
    urlcolor=cyan,
    pdftitle={Overleaf Example},
    pdfpagemode=FullScreen,
    }
\usepackage{mathrsfs}

\usepackage{times}
\usepackage{graphicx}
\usepackage{indentfirst}
\usepackage{latexsym}

\usepackage{algorithm}
\usepackage{algpseudocode}
\usepackage{amsmath}
\usepackage{physics}
\usepackage{amsfonts}
\DeclareMathOperator*{\argmax}{argmax}
\usepackage{braket}
\usepackage{multirow}
\usepackage{booktabs}
\usepackage{xcolor}
\usepackage{comment}
\usepackage{color,soul}

\usepackage{tikz}

\usetikzlibrary{tikzmark,decorations.pathreplacing}

\usepackage{tabularx}
\makeatletter
\newcommand{\multiline}[1]{%
  \begin{tabularx}{\dimexpr\linewidth-\ALG@thistlm}[t]{@{}X@{}}
    #1
  \end{tabularx}
}
\makeatother

\begin{document}

\title{Model-free Quantum Gate Design and Calibration using Deep Reinforcement Learning}

\author{Omar Shindi, Qi Yu, Parth Girdhar, and Daoyi Dong

\thanks{This work was supported by the Australian Research Council’s Future Fellowship funding scheme under Project FT220100656 and the U. S. Office of Naval Research Global under Grant N62909-19-1-2129.}

\thanks{O. Shindi, P. Girdhar and D. Dong are with the School of Engineering and Information Technology, University of New South Wales, Canberra, ACT 2600, Australia. (email: omar.shindi.ca@gmail.com, drparthgirdhar@gmail.com, daoyidong@gmail.com).}

\thanks{Q. Yu  is with the School of Engineering and Information Technology, University of New South Wales, Canberra, ACT 2600, Australia, and with Centre for Quantum Computation and Communication Technology (Australian Research Council), and also with Centre for Quantum Dynamics, Griffith University, Nathan, Queensland 4111, Australia (email: yuqivicky92@gmail.com).}}




\maketitle
\begin{sloppypar}
\begin{abstract}
High-fidelity quantum gate design is important for various quantum technologies, such as quantum computation and quantum communication. Numerous control policies for quantum gate design have been proposed given a dynamical model of the quantum system of interest. However, a quantum system is often highly sensitive to noise, and obtaining its accurate modeling can be difficult for many practical applications. Thus, the control policy based on a quantum system model may be unpractical for quantum gate design. Also, quantum measurements collapse quantum states, which makes it challenging to obtain information through measurements during the control process. In this paper, we propose a novel training framework using deep reinforcement learning for model-free quantum control. The proposed framework relies only on the measurement at the end of the control process and offers the ability to find the optimal control policy without access to quantum systems during the learning process. The effectiveness of the proposed technique is numerically demonstrated for model-free quantum gate design and quantum gate calibration using off-policy reinforcement learning algorithms.
\end{abstract}

\begin{IEEEImpStatement}
Various quantum technologies require high-fidelity quantum gate design. Many of the existing algorithms for the quantum gate design are model-based for constructing the control policy. However, a quantum system is often sensitive to noise, making it challenging to obtain precise models in many practical applications. Thus, the control policy based on quantum models will be impractical for real quantum control experiments. In this paper, we propose a novel training framework of deep reinforcement learning for model-free quantum control. The proposed framework with deep reinforcement learning treats the quantum system as a black-box and is able to find the optimal control policy by relying only on the measurements at the end of the control process. That makes the proposed method promising to be implemented in a real quantum control experiment.
\end{IEEEImpStatement}


\section{Introduction}
\IEEEPARstart{Q}{uantum} control lies at the heart of many quantum technologies \cite{Int_Gen1,Int_Gen11,Quantum_computing_gentle_introduction,Quantum Coherent Control,Q_Supremacy}. Generally, one can formalize a quantum control problem as an optimization problem and a proper control policy can then be found by minimizing the cost function related to the control goals \cite{Int_Quantum_Control,Int_Gen2,Int_Gen3,Lyapunov,Krotov,chopped_random_basis}. One major task of quantum control is quantum gate design, which aims to construct high-fidelity quantum gates. Different types of algorithms have been proposed for optimal quantum gate design \cite{Optimal_gate_design, Optimal_gate_design2}. The accuracy of the mathematical model representing the actual quantum system determines the effectiveness of the optimal control solution in real experiments. Robust quantum control increases the reliability of the control solution for experiments by encoding the noise and the experimental errors in the objective function as uncertain parameters \cite{Intro_Robust_QC, Intro_Robust_C}. Various algorithms have also been proposed for robust quantum control, with some involving machine learning \cite{R_QC1,R_QC2, R_QC3, R_QC4,R_QC5,R_QC6,R_QC7}. 

Most of the proposed techniques for quantum control are model-based methods \cite{R_QC1,R_QC2, R_QC3, R_QC4,R_QC5,R_QC6,R_QC7},\cite{RL_1, RL_2, RL_3, RL_4, RL_5, RL_6, RL_8, RL_9}, that is there is prior knowledge about the system dynamics. However, the extreme sensitivity of quantum hardware to noise makes it hard to accurately characterize quantum systems and it may not be a feasible task to derive a proper mathematical model for the effect of every influence factor in a real experiment \cite{Glassy_1,Glassy_2}. Therefore, it may be difficult to get a convincing result with model-based methods for many applications, which assume certain stability and robustness in experimental applications. Moreover, it is hard to consider all experimental constraints and errors on the model of the quantum system for robust quantum control. As a result, it leads for example to imperfect design of quantum logic gates and limited ability to reliably perform quantum computation \cite{NISQ_&_beyond}. Therefore, it is practical to suggest model-free methods in quantum control as an alternative for simulation based techniques.


 
Recently, deep Reinforcement Learning (RL) has attracted a lot of attention for application to optimal and robust quantum control \cite{RL_1, RL_2, RL_3, RL_4, RL_5, RL_6, RL_8, RL_9}, e.g., the proposed model-based deep Q-learning approach for quantum gate control \cite{RL_2}. Deep RL is a framework for machine learning algorithms that optimizes the control protocol through trial and error by studying the response of the input pulse via deep neural networks \cite{RL_introduction,DRL_QI_ER}. It is promising for model-free quantum control, due to its ability to identify strategies for achieving a goal in a complex space of solutions without prior knowledge of the system \cite{DRL_Survey_2017}. For example, a circuit-based RL approach has been proposed for model-free quantum state preparation \cite{Free_RL}.

In this paper, we propose a training framework for deep RL for model-free quantum control with limited control resources. To illustrate the effectiveness of our framework through numerical examples, we consider the quantum gate design problem in a model-free way for three different cases. The first case is a problem where a matrix representing the quantum gate is expected at the end of the control process. The goal is to obtain a designed gate close, with respect to the gate fidelity, to the desired quantum gate. The second and third cases involve quantum gate calibration, similar to quantum Hamiltonian tomography \cite{Tomo_1, Tomo_2}, where a quantum state is utilised to assess the effectiveness of the calibrated gate. The general idea is to perform a calibrated quantum operator on a variety of quantum states, and then compare the results to the desired states. For the second case, a single qubit gate is calibrated, which performs an operation composed of a series of single qubit gates that are part of a certain universal gate-set. For the third case, the quantum gate calibration is within a quantum circuit. The simulation results demonstrate the effectiveness of the proposed approach for designing and calibrating quantum gates with limited control resources. Also there is potential for the proposed approach to be applied in real quantum experiments without the need to access the quantum state during the training. The main contributions of this work can be summarized as follows:
\begin{itemize}
    \item {Proposing a novel model-free quantum control framework with deep RL that treats the quantum system as black-box}.
    \item {Adopting the proposed model-free quantum control framework for achieving quantum gate design task}.
    \item {Considering the problem of quantum gate calibration within a quantum circuit, and employing the proposed model-free method for solving this problem}.
\end{itemize}

The rest of the paper is organized as follows. Preliminaries of quantum gate design are explained in Section \ref{Related_Work}. Section \ref{RL_Backgruond} briefly introduces RL. The proposed RL framework for model-free quantum control is explained in Section \ref{Proposed_Approach}. In section \ref{results} the performance of the proposed framework is illustrated through simulation results. Finally, Section \ref{Conclusion} draws out the conclusions.



\section{Quantum gate design}\label{Related_Work}

High-fidelity quantum gate design is critical for the success of quantum technology applications like quantum computation and quantum communication. A quantum gate mathematically can be represented by a unitary matrix $U$ of size $2^d \text{x} 2^d$ in complex Hilbert space $\mathscr{H}$, where $d$ is the number of qubits that the quantum gate is acting upon. The unitary operator $U$ is used to transform an initial state $\ket{\psi_0}$ to a desired state $\ket{\psi_T}=U\ket{\psi_0}$. Practically, quantum gates are often approximated using a sequence of control pulses $\{A_1,...,A_N\}$, with a constant pulse duration $dt = T/N$, where $T$ is the total evolution time and $N$ is the total number of control pulses. The main task of the quantum gate design problem is to find the right control protocol that can approximate the unitary operator $U_f = U(A_N)\leftarrow U(A_{N-1})...\leftarrow U(A_1) \leftarrow U_0$, beginning from the initial unitary $U_0$, to the desired unitary $U_T$. For model-based quantum control methods, the Hamiltonian $H$ of the quantum system is given. Thus, the quantum gate $U(A_t)$ for the applied control pulse $A_t$ at time step $t$ can be found by using the Schr\"{o}dinger equation as  
\begin{equation}\label{Uj}
U(v_t)=e^{-iH(A_t)dt} U(A_{t-1}),
\end{equation}
where \(i\) is unit imaginary number, $H(A_t)$ is the Hamiltonian of the quantum system, $U(A_{t-1})$ is the unitary operator at the previous time step $t-1$. The quality of the approximated gate $U_f$ with respect to the desired gate $U_T$ can be checked by computing the fidelity $F_f$ \cite{RL_3,Fidelity_1} as

\begin{equation}\label{Final Fidleity_quantum circuit 1}
\centering
\displaystyle 
F_f =\left|\frac{\text{Tr}[U_{f}^\dagger U_T ]}{2^d}\right|^2,
\end{equation} 
and the goal is to approximate the quantum gate with high fidelity. Here $\text{Tr}[X]$ returns the trace of $X$ and $X^{\dagger}$ represents the transpose and conjugate of $X$. 

\section{Reinforcement learning background}\label{RL_Backgruond}
Reinforcement Learning (RL) is a machine learning technique, in which an agent, or multi-agents, learns to do a specific task or tasks by trial and error via interacting with the environment \cite{RL_introduction}. The agent interaction cannot change the dynamics or rules of the environment, which represents the problem that the agent is trying to solve. At the learning stage, the RL agent interacts with the environment on a series of episodes. On each episode, the RL agent interacts with the environment in a sequence of discrete time steps $t=1,2,3..,N$ with fixed duration $dt$. At time step $t$ the environment provides the RL agent a state observation $S_t$ that describes the system at time $t$. The RL agent responds by selecting an action $A_t$, which yields the next state $S_{t+1}$ after evolution. Then, the quality of the applied action $A_t$ for achieving the control goal can be indicated by the return reward $R_t$. The ultimate goal of the RL agent is to maximize the return rewards. The episode will be terminated if one or more of termination conditions have been satisfied like reaching the maximum number of time steps $N$. Here we briefly introduce three deep RL algorithms including Deep Q-learning (DQL), double DQL and dueling that are used in our quantum control tasks.

\subsubsection{Deep Q-learning (DQL)}

Deep Q-learning  is a value-based RL algorithm using Neural Networks (NNs) to approximate Q-values that represent the expected future returns for action-state pairs as a replacement for tabular representation. Thus, the DQL method is able to solve more complex or high-dimensional problems \cite{DRL_Survey_2017,DQL_2015}. Generally, a DQL agent contains two neural networks of the same architecture: the value-network with weights \(\theta_V\), and target-network with weights \(\theta_T\). The value-network receives current state \(S_t\) and returns Q-values \(Q(S_t,\mathcal{A},\theta_V)\) for all allowed actions $a_1,a_2...,a_p$ in the action space \(\mathcal{A} \in [a_1,a_2...,a_p]\). The target-network receives next-state \(S_{t+1}\) and returns Q-values \(Q(S_{t+1},\mathcal{A},\theta_T)\) for all actions. At instant time \(t\) the DQL agent chooses the action \(A_t\) based on a specified procedure, like the epsilon-greedy method,
{ \begin{equation} \label{Greedy_Process}
  \centering
    A_t=
    \begin{cases}
        \argmax\limits_{a}\{Q(S_t,\mathcal{A},\theta_V)\},& x < \epsilon, \\
        \text{a random action} \in \mathcal{A}, & \text{otherwise},\\
    \end{cases}
  \end{equation}}where \(\epsilon \in [0,1]\) is epsilon-greedy parameter, and \( x \in [0,1 ]\) is chosen randomly to achieve balance between exploitation and exploration for action selection from action space \(\mathcal{A}\). Mainly, the goal is to construct the optimal control protocol $A^*=[A_1,A_2,...,A_N]$ with a high chance of achieving the optimised problem objective. To accomplish this, the DQL agent would determine the optimal Q-function $Q^*$, which generates the maximum cumulative discounted rewards at the end of each episode,
 \begin{equation} \label{Goal_of_DQL} 
 \centering
Q^*=\argmax\limits_{\theta_V}\sum_{t=1}^{N} Q(S_t,A_t,\theta_V).
\end{equation}
State-transition experience \(E_j=\{S_t,A_t,R_t,S_{t+1}\}\) will be stored at replay experience memory \(Me=\{E_1,E_2,...E_b\}\) with size \(b\) for later use of selecting randomly Mini-batch samples \(Mb_{samples}\) with size \(K\) to train the value network. 

The target-network is required for supervised learning to compute target-value or expected maximum Q-value \(\max_{A}\{Q(S_{t+1},A,\theta_T)\}\) at next-state \(S_{t+1}\) for each sample of \(Mb_{samples}\) by applying the following Q-learning update
  \begin{equation} \label{target_value}
  \centering
    Q_T = R_t+\gamma(\max_{A}\{Q(S_{t+1},A,\theta_T)\}),
  \end{equation}
where $\gamma$ is the discount reward factor. Then, the Mean Square Error (MSE) is adopted to evaluate loss between predicted and target Q-values
\begin{equation} \label{MSE}
\centering
l = \text{MSE}(Q(S_t,A,\theta_V) - Q_T).
\end{equation}
Parameter \(\theta_V\) of the value-network  will be updated to minimise the loss value \(l\) by using a Gradient Descent (GD) optimizer with learning rate \(\alpha\)
\begin{equation}\label{GD}
\centering
\theta_{V+1} \leftarrow \theta_V - \alpha(\nabla_{\theta_V}l|_{\theta_V}),
\end{equation}
where \(\nabla_{\theta_V}l|_{\theta_V}\) is the gradient of loss with respect to \(\theta_V\). However, weights of the target-network will be updated as \(\theta_T \xleftarrow[]{} \theta_V\) every \(Z\) episodes to be equal to the weights of the value network $\theta_V$. The learning procedure for DQL agent keeps repeating until any of the termination conditions, like the maximum number of episodes, is achieved. At the end of training, DQL agent is expected to converge to the optimal control policy.

\subsubsection{Double DQL}
The standard DQL may suffer from overestimation due to using the same value of the max operator for action selection in (\ref{Greedy_Process}) and in (\ref{target_value}) for action evaluation. To solve this issue and to reduce the overestimation in the loss function, the Double Q-learning \cite{Double_DQL} has been proposed to use two sets of weights $\theta_T$ and $\theta_{T}^{'}$ for the action evaluation, 
  \begin{equation} \label{Double_DQL}
  \centering
    Q_T = R_t+\gamma Q(S_{t+1},\argmax\limits_{A}\{Q(S_{t+1},A,\theta_T)\},\theta_{T}^{'}).
  \end{equation}

\subsubsection{Dueling Network}
The dueling network is a single Q-network architecture, using two streams of fully connected layers to estimate the state value $V(S)$ and the advantage of each action $Q^{'}(S,A)$ \cite{Dueling_Double_DQL}. The Q-values for the actions $A \in \mathcal{A}$ at the state $S$ can be computed as,
  \begin{equation} \label{Dueling_Q_value}
  \centering
    Q(S,A)= V(S)+Q^{'}(S,A).
  \end{equation}
The dueling architecture helps to converge faster than standard DQL. The dueling network is usually applicable only for value-based RL algorithms \cite{Rainbow}.


\section{Model-free quantum gate design and quantum gate calibration}

In contrast to the model-based quantum gate design explained in Section \ref{Related_Work}, the dynamic model of the quantum system is not available in the model-free case. Thus, the RL agent is not aware of the approximated model of the quantum system, and the RL agent is dealing with the quantum system as a black box. In this paper we consider the quantum gate design problem in a model-free way for three different scenarios as follows.

\subsection{Model-free quantum gate design}\label{Free_model_gate_design}
\begin{figure}[H]
\centering
\includegraphics[width=8 cm,height=3.5cm]{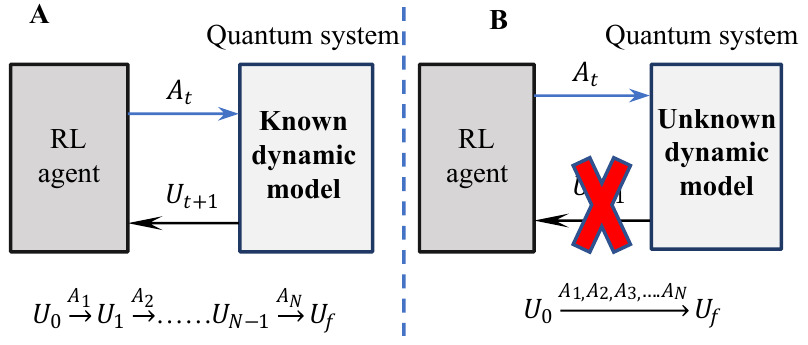}
\caption{(A) Model-based RL approach for quantum gate design, (B) RL for model-free quantum gate design. } \label{Fig:Figure 1}
\end{figure}

The goal of quantum gate design is to generate a quantum gate, using available operations, to perform a desired operation on a quantum system. The RL agent algorithm is supposed to be a useful and effective method for finding such a proper control protocol $A^*$ which can approximate the desired unitary gate $U_T$. The main difference between model-based and model-free quantum gate design is given in Figure\ref{Fig:Figure 1}. In Figure \ref{Fig:Figure 1}(A), the dynamics of the quantum system is used to provide the observation $U_t$ as feedback to the RL agent after receiving the control action $A_t$, and then $U_t$ will be used by the RL agent to choose the next action $A_{t+1}$.

Figure \ref{Fig:Figure 1}(B) explains model-free quantum gate design. As shown in Figure \ref{Fig:Figure 1}(B), the feedback of information $U_t$ is not available during the evolution process for choosing the next action. In this scenario that we consider, the RL agent performs the sequence of the control protocol, and it is assumed that $U_f$ can be directly retrieved. The goal is to approximate $U_f$ to the desired quantum operator. In this case, the fidelity of the approximated quantum operator $U_f$ can be computed by using (\ref{Final Fidleity_quantum circuit 1}), to be used later for computing the reward.


\subsection{Model-free calibration of a composed single-qubit gate}\label{Calibration_gate2}
\begin{figure}[H]
\centering
\includegraphics[width=8.0 cm,height=5 cm]{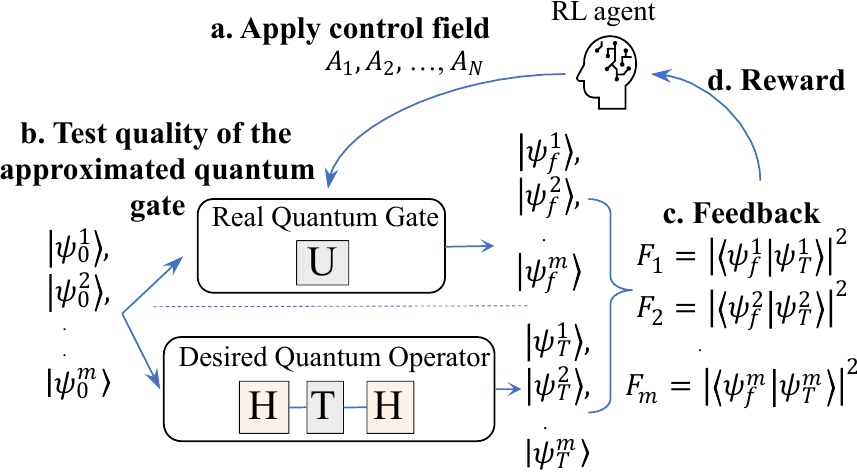}
\caption{Procedure for model-free quantum gate calibration of a composed single-qubit gate.}
\label{Fig:Calibration of single gate}
\end{figure}

In this case, we consider the calibration of the quantum gate for a single-qubit system whose dynamic model is unknown to the RL agent. A quantum operations is formed from a sequence of a set of available quantum gates like Hadamard and Pauli gates to get the desired change on the quantum state. The goal in this case is to optimize a single quantum gate to the desired operator that requires a sequence of certain single-qubit gates to be implemented. Figure \ref{Fig:Calibration of single gate} explains the training sequence of the RL approach for the model-free calibration of composed single qubit gate $\text{U}$.

As shown in Figure \ref{Fig:Calibration of single gate}, the RL agent will apply a sequence of control pulses $A_1,A_2,..,A_N \in \mathcal{A}$ to calibrate $\text{U}$. The calibrated quantum gate will next be evaluated by feeding a set of training states $\{\ket{\psi_{0}^{1}}, \ket{\psi_{0}^{2}}..., \ket{\psi_{0}^{m}}\}$ as inputs to the quantum gate. Then, the fidelity of the output states $\{\ket{\psi_{f}^{1}}, \ket{\psi_{f}^{2}}..., \ket{\psi_{f}^{m}}\}$ with respect to the target quantum states $\{\ket{\psi_{T}^{1}}, \ket{\psi_{T}^{2}}..., \ket{\psi_{T}^{m}}\}$ is computed as
\begin{equation}\label{Final Fidleity_quantum circuit}
\centering
F_j =|\braket{\psi_{f}^{j}| \psi_{T}^{j}} |^2, j=1,2,...,m.
\end{equation}
As result of the quality testing process for the approximated operator, we will have a vector of fidelities $\vec{F}=\{F_1,F_2,....,F_m\}$ that describes the quality of the calibrated single quantum gate for each training set. The control objective of the RL agent is to calibrate the composed single-qubit gate $\text{U}$ to make the worst fidelity $\min(\vec{F})$ as high as possible.



\subsection{Model-free quantum gate calibration within quantum circuit}\label{Calibration_gate1}
\begin{figure}[H]
\centering
\includegraphics[width=8.3cm,height= 5.2 cm]{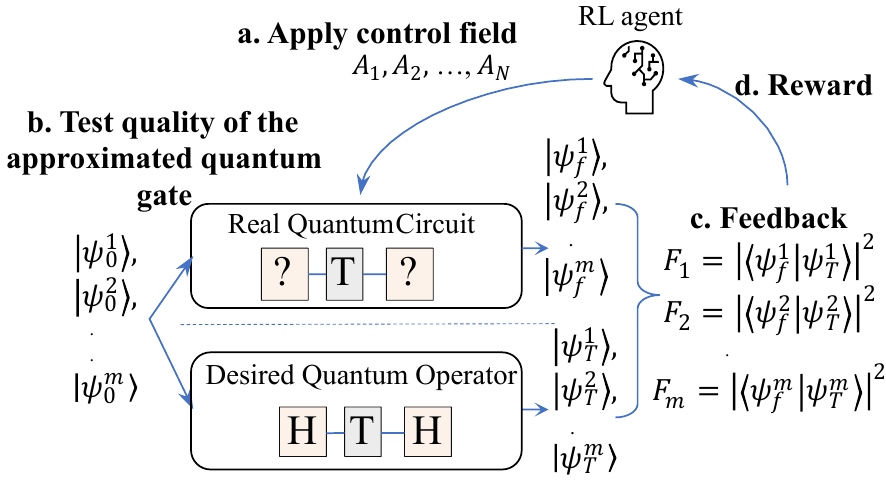}
\caption{Procedure for model-free quantum gate calibration within quantum circuit.} \label{Fig:Gate_Calibration}
\end{figure}

In this case,  we consider the problem of tuning and calibrating the quantum gate within a quantum circuit whose model is unknown to the employed RL agent. The RL agent is interacting with the target quantum system as a black box. Figure \ref{Fig:Gate_Calibration} explains the training process of the RL approach for the model-free quantum gate calibration within quantum circuit.

As shown in Figure \ref{Fig:Gate_Calibration}, the RL agent first applies control pulses $A_1,A_2,..,A_N \in \mathcal{A}$ from the action space $\mathcal{A}$ to calibrate the specified quantum gates inside the quantum circuit. Then, the calibrated quantum gates will be evaluated by performing on a variety of quantum states $\{\ket{\psi_{0}^{1}}, \ket{\psi_{0}^{2}}..., \ket{\psi_{0}^{m}}\}$ to the quantum circuit. The output results $\{\ket{\psi_{f}^{1}}, \ket{\psi_{f}^{2}}..., \ket{\psi_{f}^{m}}\}$ will next be compared to the target states $\{\ket{\psi_{T}^{1}}, \ket{\psi_{T}^{2}}..., \ket{\psi_{T}^{m}}\}$ using fidelity calculated as in (\ref{Final Fidleity_quantum circuit}). As a result, a vector of fidelities $\vec{F}=\{F_1,F_2,....,F_m\}$ that describes the quality of the calibrated quantum gate for each training set will be produced. The vector of fidelities $\vec{F}$ will next be provided to the RL agent for learning. The objective is to calibrate the quantum gate within the quantum circuit to make the worst fidelity $\min(\vec{F})$ as high as possible.



\section{Reinforcement learning for model-free quantum gate design and calibration}\label{Proposed_Approach}

Most existing RL approaches for quantum control problems are model-based \cite{RL_1}-\cite{RL_9}. For example a Dueling Double DQL approach has been proposed for model-based quantum gate design \cite{RL_2}, which supposes knowing the quantum operator $U_t$ after each control pulse. The $U_t$ will be used by the RL agent to choose the next action. In the model-free approach as explained in Figures \ref{Fig:Figure 1}(B), \ref{Fig:Calibration of single gate} and \ref{Fig:Gate_Calibration}, the RL during the evolution process has no information about the quantum unitary. In our procedure, we construct the state $S_t$ from the available information as follows. At each time step $t \in [1,2,..N]$ the state $S_t$ is equal to $S_t=[A_t/z, (t-1)/N]$, whereas $A_t=[u_{0}^t,u_{1}^t,...,u_{n}^t]$ is the control vector containing the values of the control fields $u_0,u_1...u_n$ at the time step $t$, while $z$ is a normalization factor. Simply, the proposed framework will allow the DQL agent to use the applied action $A_t$ and the time step $t$ to choose the suitable next action $A_{t+1}$. Algorithm \ref{New Procedure} explains the proposed procedure. As shown in Algorithm \ref{New Procedure}, for every episode the RL agent will start with constructing the control protocol ${A_1,A_2,...,A_{N}}$, then performing the control protocol on the quantum system. Finally, the reward based on the measurement at final time can be obtained.



\subsection{Action selection method}

The $\epsilon$-greedy method is used for the action selection process. The first action will open the evolution process, and a good choice will lead to good results and vice versa. Hence, it is important to make a good choice for the first action before using the prediction network to choose the rest of the actions. The first action $A_1$ will be chosen without using the prediction network as follows, 

{ \begin{equation} \label{First_action}
  \centering
    A_1=
    \begin{cases}
        A_1^{Best},& x < \epsilon, \\
        \text{a random action} \in \mathcal{A}, & \text{otherwise}.\\
    \end{cases}
  \end{equation}}
In the case of exploitation, the action of the best discovered experiences $A_1^{Best}$ will be used, and otherwise the action will be chosen randomly. The rest actions $A_2,A_3,...,A_N$ will be selected as explained in (\ref{Greedy_Process}). The value of \(\epsilon\) as shown in (\ref{Greedy_Process}) defines the percentage of exploitation and exploration, to prevent RL agent from sticking to the local optimal results and to assist the RL agent to approach the global optimal results. The value of epsilon \(\epsilon\) will be updated after each episode by adding $\epsilon_{step}$, until it reaches the maximum value $\epsilon_{max}$.

{ \begin{equation} \label{epsilon_update}
  \centering
    \epsilon=
    \begin{cases}
        \epsilon+\epsilon_{step}, & \epsilon < \epsilon_{max},\\
        \epsilon_{max}, & \text{otherwise}.\\
    \end{cases}
  \end{equation}}

\begin{algorithm}[h]
\caption{DQL Training Procedure for Model-free Quantum Gate Design}\label{New Procedure}
\begin{flushleft}
{\textbf{Input:} {\text{Evolution time} $T$}, {\text{Number of episodes} $N_e$}, {\text{Actions space} $\mathcal{A}$}, {\text{Normalization parameter} $z$}, {\text{Control steps} $N$}, {\text {Final exploration - exploitation percentage} $\epsilon_{max}$}, {\text {Learning rate} $\alpha$}, {\text{Experience memory size} $b$}, {\text {Reward discount} $\gamma$}, {\text{Size of training mini-batch} $K$}, {\text{Training predict weights} $n$}, {\text{Replacement target weights} $Z$}, {\text{Storing Best Experience} $k$}.}
\end{flushleft}

\begin{flushleft}
{\textbf{Pre-process:} {\text{Pulse duration} $dt = T/N$}, {\text {Best fidelity} $F_{Best}=0$}, {\text{Epsilon } $\epsilon = 0 $}, {\text{ Epsilon step} $\epsilon_{step}$},{\text{ Total control steps} $Step=0$}.}
\end{flushleft}

\begin{algorithmic}[1]
    \For{\text{e=1,2,....,$N_e$}}\Comment{Loop for episodes}
    \State{Clear the episode buffer $E=[]$.}
    \State {Initialize unitary operator to $U_0$.}
    \State{Choose the first action $A_1$ according to Eq. (\ref{First_action}).}
    \State{\text{Construct the state} $S_{1}=[A_1/z, 0]$.}
    \For{\text{i=2,3,....,$N$}}\Comment{Loop for control steps}
    \State{Choose the action $A_\text{i}$ according to Eq. (\ref{Greedy_Process}).}
    \State{\text{Construct the state} $S_{\text{i}}=[A_\text{i}/z, (\text{i}-1)/(N)]$}.
    \State{$E_\text{i} \gets \{S_{\text{i}-1}, A_{\text{i}-1}, S_\text{i}\}$}
    \State{$Step \text{ += } 1 $} \Comment{Control steps counter}
    \If {\(mod(Step,n)==0\)} \Comment{Every \(n\) Steps}
        \State{Update the value network.}
    \EndIf
    \EndFor 
    \State{Apply the control protocol to the quantum system.}
    \State {$A_1 \xrightarrow[]{\delta_t} A_2 \xrightarrow[]{\delta_t} A_3 ..... \xrightarrow[]{\delta_t} A_{N}$}
    \State{Get final unitary $U_f$.}
    \State{Compute the fidelity $F$.}
    \State{Compute the reward  $ R =-log(1-F)$.}
    \State{Reward the episode experience $E$ by $R$.}
    \State  \multiline{%
    Store the episode experience (${[E_1,R],...,[E_{N},R]}$) \\ 
    into the replay experience memory (REM).}
    \If{($F > F_{Best}$)}
           \State {$F_{Best}=F$}\Comment{Store best fidelity }
          \State{Store the episode experience $E_{Best}=E$.}
    \EndIf

    \If {\(mod(e,k)==0\)} \Comment{Every \(k\) episodes}
    \State {Store best episode experience $E_{Best}$ into REM.}
    \EndIf
    \If {\(mod(e,Z)==0\)} \Comment{Every \(Z\) episodes}
    \State{Update the Target network.}
    \EndIf

\EndFor
\end{algorithmic}
\end{algorithm}

\subsection{Rewards and Modified Experience Memory}

In general, the goal for solving the quantum control problem is to find a proper control sequence that steers the quantum system from the initial unitary  $U_0$ to the desired target operator $U_T$. The ultimate goal of the RL agent is to maximize the collecting reward \(R_T\) that defines the quality of the applied control protocol. In our framework, the RL agent will receive the reward $R$ at the end of each episode and after performing the control protocol. This reward value is dependent on the final fidelity $F_f$, 

\begin{equation}\label{Rewards_quantum_gate_design}
\centering
R=-log(1-F_f).
\end{equation}

The episode transition experience will be stored into a buffer $E=\{[S_1,A_1,S_2],...,[S_{N},A_{N},S_{N+1}]\}$, then a $n$-step delay reward function will be applied to give all the episode state transitions the same reward at the end of each episode based on final unitary of the quantum system  \(U_f\). The fidelity for the quantum gate design problem can be calculated using (\ref{Final Fidleity_quantum circuit 1}). 

The goal for quantum gate calibration as explained in Sections \ref{Calibration_gate1} and \ref {Calibration_gate2} is calibrating the quantum system to get the worst fidelity $min(\vec{F})$ of the training set with highest fidelity as possible. Thus the reward for the quantum gate calibration problem can be computed as 

\begin{equation}\label{Rewards_Quantum_gate_calibration}
\centering
R=-log(1-\min(\vec{F})).
\end{equation}
Then, the episode state transition with the same reward $\{[E_1,R],[E_2, R]...,[E_N,R]\}$ will be saved to the experience reply memory $M_e$ to be used later to train the RL agent. This rewarding method will keep all the episodes of transition states linked to each other, and make changes on the weights of the prediction network. This will improve the ability for the RL agent to distinguish the difference between the different inputs and make it more likely to find better results. During the training, we keep monitoring achieved final fidelity for each episode and store the transitions of the one with the highest fidelity, to be used for the action selection of the first action $A_1$. To increase the performance of the RL-agent to find better results, the best discovered experience will be added to the experience replay memory frequently every specified number of episodes, to increase the chance of using it for the training.

\section{Results and Discussion}\label{results}

The proposed framework has been implemented with four DQL algorithms, the Model-free DQL (MDQL), Model-free Double DQL (MDDQL), Model-free Dueling DQL (MDuDQL) and Model-free Double Dueling DQL (MDuDDQL). The performance of MDQL, MDDQL, MDuDQL and MDuDDQL is tested for the quantum gate design problem for single and two qubit systems. They are also applied to a single gate and quantum gate calibration of Hadamard and CNOT gates within a quantum circuit.

The results in this paper are generated on a workstation computer with a dual processor Intel(R)Xeon(R)W-1245, 64 GB RAM. The algorithm is implemented in Python. Codes for model-free quantum gate design and calibration using RL associated with the current submission are available at \href{https://github.com/Omarshindi/Model-Free-quantum-gate-design-and-calibration-using-Deep-Reinforcement-Learning.git}{GitHub}\footnote{https://github.com/Omarshindi/Model-Free-quantum-gate-design-and-calibration-using-Deep-Reinforcement-Learning}. The following table lists the main parameters used in the simulations.

\begin{table}[H]
\footnotesize
\centering
\caption{Parameter values of various algorithms}\label{table_1}
\begin{tabular}{|l|l|}
 \hline
\textbf {Parameter} & \textbf{Value} \\  \hline

  Learning Rate $(\alpha)$ &  0.005 \\\hline 
  Reward Discount $(\gamma)$ & 0.95 \\ \hline
  Number of Episodes (E)&  \(2 * 10^5\) \\ \hline
  Size of Hidden-Layer & 512 \\ \hline
  Experience Memory Size (\(b\)) &  25000     \\ \hline        
  Size of Mini-batch (\(K\)) &  64   \\ \hline
  Training Predict Weights (\(n\)) & 10 (Time steps) \\ \hline
  Replacement Target Weights (\(Z\))& 10 (Episodes) \\ \hline
  Epsilon Updating Step \(\epsilon_{step}\) & 0.0001 \\ \hline 
  Normalization Parameter (\(z\)) & 40 \\ \hline 
  Storing Best Experience (\(k\)) & 3 (Episodes)\\ \hline
 \end{tabular}
 \end{table}

The value of $\epsilon_{max}$ is defined as follows based on the value of the final fidelity $F_f$, 
{ \begin{equation} \label{epsilon_max_update}
  \centering
    \epsilon_{max}=
    \begin{cases}
        0.9999, & 0.99 \leq F_f < 0.999,\\
        0.99999, & 0.999 \leq F_f,\\
        0.95, & \text{otherwise}.\\
    \end{cases}
  \end{equation}}

The following Hamiltonian has been used for the single qubit system,

\begin{equation}\label{Hamiltonian_H}
    H=u_{0}\sigma_{z}+u_{1}\sigma_{x}
\end{equation}
with control fields $u_{0},u_{1}\in \{-4,4\}$; each control field can take one value of two allowed actions. And the $\sigma_{j} \in \{\sigma_{x},\sigma_{y},\sigma_{z}\}$ are standard Pauli operators.

For the 2-qubit system, the following Hamiltonian has been used in the simulator,

\begin{equation}\label{Hamiltonian_CNOT}
    H=S_z+u_0 S_{x}^{1}+u_1 S_{x}^2+u_2 S_{y}^1+u_3 S_{y}^{2}
\end{equation}
whereas, 
\begin{equation}\label{Sx}
    S_{z} = \sigma_{z}\otimes\sigma_{z},
\end{equation}
\begin{equation}\label{Sx1}
    S_{j}^{1} = \sigma_{j}\otimes\mathbb{I},
    S_{j}^{2} = \mathbb{I}\otimes\sigma_{j}.
\end{equation}
Here $\mathbb{I}$ is the identity matrix with size 2 x 2  and the operation $\otimes$ denotes tensor product. Each control field $u_0,u_1,u_2,u_3 \in [-4,4]$ can take one value of two allowed actions. We emphasize again that the proposed model-free control approach is dealing with the quantum system as a black-box and choose these specific quantum systems to compare the performance of the proposed  model-free quantum gate design approach with an existing model-based RL approach.

\subsection{Results for quantum gate design}
The goal of the quantum gate design problem is finding the proper control protocol that can steer the applied unitary on the quantum system from $U_0$ to $U_f$ at the end of the evolution process as close as possible to the target gate $U_T$. 

\subsubsection{Hadamard gate}\label{section 1}

The Hadamard gate is an important operation for quantum computation. As mentioned earlier the DQL agent does not have any access to the quantum system to get the quantum state after each control step. For the simulator of the quantum system and for the comparison with an existing RL approach for model-based quantum gate design in \cite{RL_2}, we have created a simulator for single-qubit quantum gate shown in (\ref{Hamiltonian_H}) and considered $u_0=1$ all the time. The final evolution time is equal to $T=1$. The DQL agent will interact over the total number of discrete steps $N=28$, and the effective time for each step is equal to $\delta=T/N$. In this case, the RL agent interacts with an environment representing the quantum gate system as explained in Section \ref{Proposed_Approach}. The initial unitary on the quantum system is considered as the identity matrix of size 2x2 and the target is the Hadamard quantum gate. The infidelity has been utilised to assess the quality of the approximate unitary $U_f$. Mathematically, the infidelity is equal to $1-F_f$ where the fidelity $F_f$ is calculated by using (\ref{Final Fidleity_quantum circuit 1}). The infidelities of the best designed gate for four algorithms are listed in Table \ref{Infid_Hadamard}.

\begin{table}[H]
\footnotesize
\centering
\caption{The best achieved infidelity for Hadamard quantum gate design by using MDQL, MDDQL, MDuDQL, and MDuDDQL.}\label{Infid_Hadamard}
\begin{tabular}{llll}
\hline
Algorithm & & & Infidelity \\ \hline
MDQL      & & & 0.00021   \\
MDDQL     & & & 0.00006    \\
MDuDQL    & & & 0.00026    \\
MDuDDQL   & & & 0.00008    \\ \hline 
\end{tabular}
\end{table}

The MDDQL and MDuDDQL have constructed the gate of the lowest infidelity less than $10^{-5}$, followed by the MDQL and MDuDQL that achieve a little over $10^{-4}$. The RL agent of model-based quantum gate design as shown in results in \cite{RL_2} achieved around $10^{-4}$ for Hadamard gate design under the same settings. The RL agent of MDDQL and MDuDDQL are able to find good results, similar to those obtained using the model-based framework in \cite{RL_2} for Hadamard gate design under the same quantum model and control parameters.

Figure \ref{Fig:Figure 2} below shows the average achieved infidelity vs the number of episodes for MDQL, MDDQL, MDuDQL and MDuDDQL for the Hadamard quantum gate design. 

\begin{figure}[H]
\includegraphics[width=9.2 cm,height= 8.0 cm]{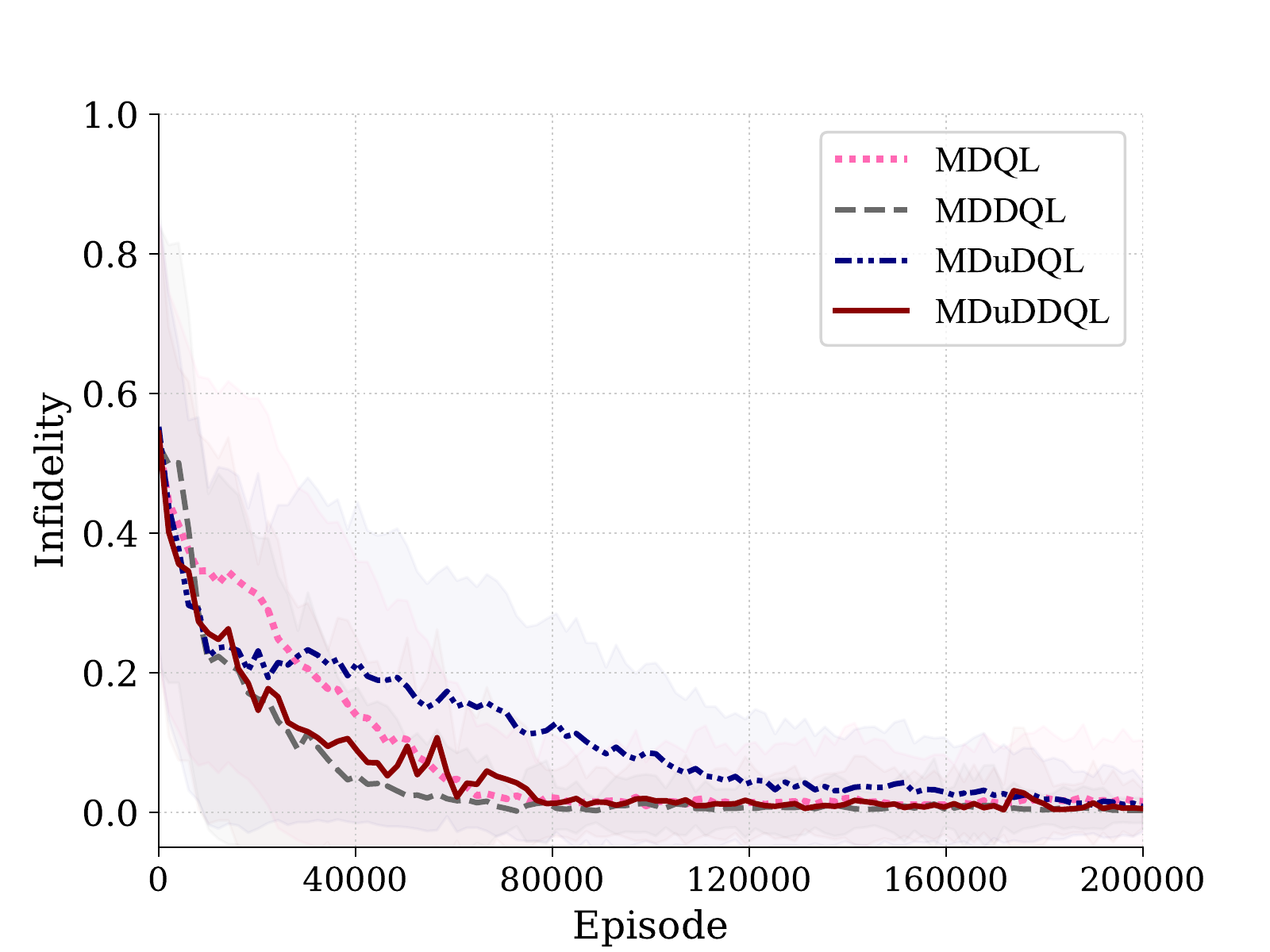}
\caption{The solid lines represent the average achieved infidelity of 2000 samples during the training for the Hadamard gate design problem while the highlighted area represents the standard deviation.} \label{Fig:Figure 2}
\end{figure}

MDQL, MDDQL, MDuDQL and MDuDDQL as shown in Figure \ref{Fig:Figure 2} converge to a low infidelity of control policy that can construct a high fidelity of control protocol without any access to the quantum state during the control process. 


\subsubsection{CNOT gate} \label{section 2}

Controlled-NOT or CNOT gate, a two qubit quantum gate, is one of the essential quantum gates for quantum computation and communication. For the simulator of the quantum system and for the comparison with an existing RL approach for model-based quantum gate design in \cite{RL_2}, we have created a simulator for two-qubit quantum gate shown in (\ref{Hamiltonian_CNOT}). The goal is to find a proper control protocol with the number of steps $N=38$, and pulse duration $\delta=1.1/38$ to get at the end of evolution process the final gate close to the CNOT gate. Table \ref{Infid_CNOT} contains the infidelities of the best designed gate to CNOT gate for four algorithms.

\begin{table}[H]
\footnotesize
\centering
\caption{The best achieved infidelity for CNOT quantum gate design by using MDQL, MDDQL, MDuDQL, and MDuDDQL.}\label{Infid_CNOT}
\begin{tabular}{llll}
\hline
Algorithm & & & Infidelity \\ \hline
MDQL      & & & 0.0556     \\ 
MDDQL     & & & 0.0096     \\ 
MDuDQL    & & & 0.0841     \\ 
MDuDDQL   & & & 0.0089     \\ \hline
\end{tabular}
\end{table}

\begin{figure}[H]
\centering
\includegraphics[width=9.3 cm,height=7.9 cm]{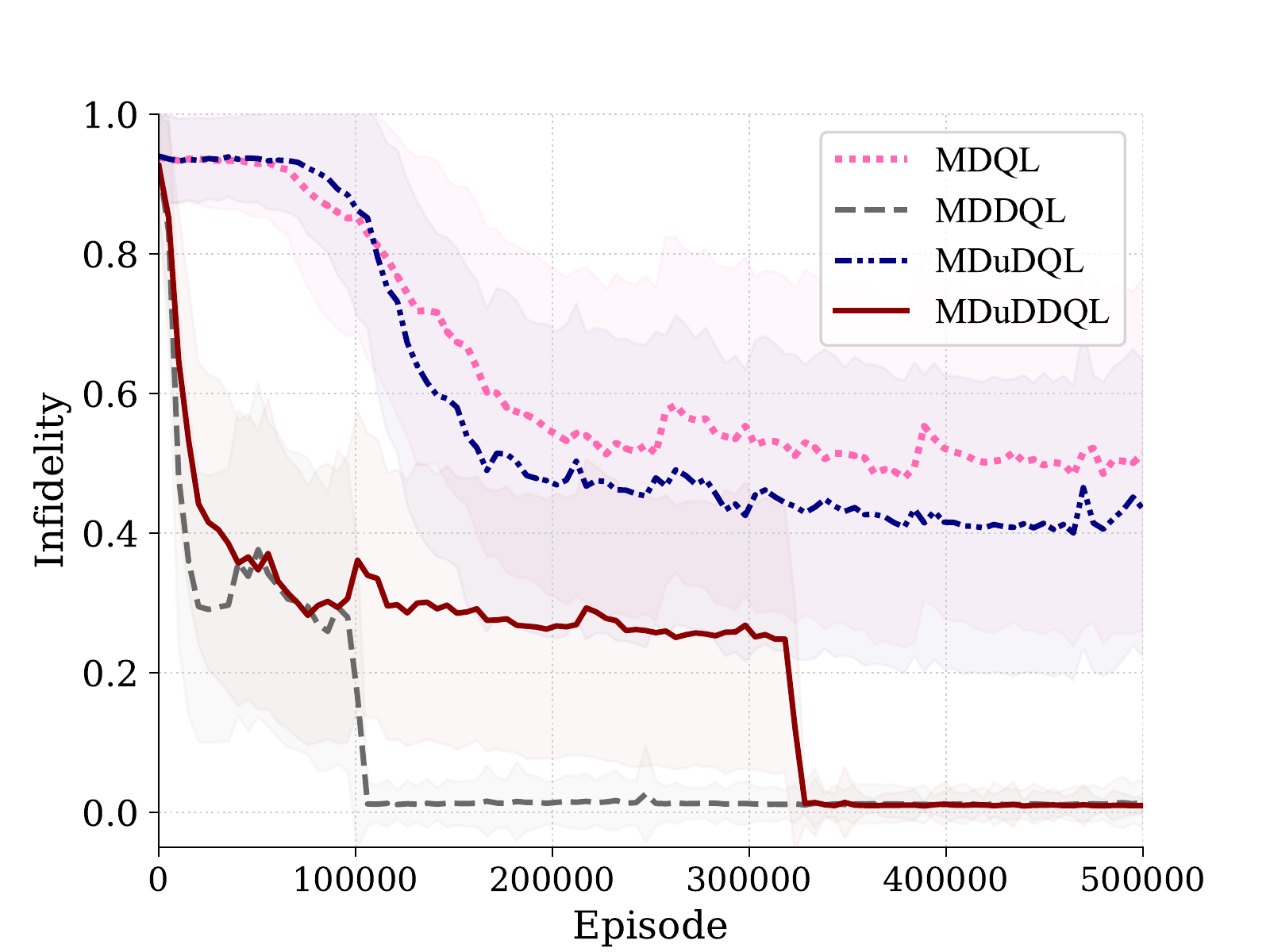}
\centering
\caption{The solid lines represent the average achieved infidelity of 5000 samples during the training for the CNOT gate design problem and the highlighted area represents the standard deviation.} \label{Fig:Figure 3}
\end{figure}

As shown in Table \ref{Infid_CNOT}, the MDDQL and MDuDDQL succeed to find a high fidelity of quantum gate with infidelity less than $10^{-2}$ while MDQL and MDuDQL have failed to achieve this level. MDQL and MDuDQL are using single Q-value estimation function that causes an overestimation for the Q-values. This overestimation harms the performance and causes getting stuck into local optimal solutions. Figure \ref{Fig:Figure 3} shows the average achieved infidelity for the CNOT gate design problem vs the number of episodes for MDQL, MDDQL, MDuDQL and MDuDDQL algorithms.

As shown in Figure \ref{Fig:Figure 3}, MDQL and MDuDQL have failed to construct a high-fidelity control protocol and they have gotten stuck to local optimal solutions. However, the MDDQL and MDuDDQL agents with the proposed framework have succeeded to converge to a high-fidelity control policy that can achieve infidelity less than $10^{-2}$. The success of MDDQL and MDuDDQL is attributed to the usage of the double DQL, which reduces Q-value overestimation and allows for more reliable learning and discovery of better solutions during training. The jump in the performance of MDuDDQL and MDDQL may be due to the change in the value of $\epsilon_{max}$ as explained in (\ref{epsilon_max_update}).

In \cite{RL_2}, standard DuDDQL was used for CNOT model-based quantum gate design and the best achievable infidelity was around $10^{-3}$. The proposed Model-free RL performs as well for designing quantum gates as model-based RL when compared to the results in \cite{RL_2}.


\subsection{Composed single-qubit gate} \label{U_single_gates}

Here we consider two examples of gates that are equivalent to a series of certain gates which are part of a typical universal gate-set. The first one is calibrating the T$_{\text{x}}$ operation that rotates the qubit around the x-axis by 45 degrees. The second example is designing the T$_{\text{y}}$ that rotates the qubit around the y-axis by 45 degrees.

\subsubsection{T$_{\text{x}}$ gate design}

\begin{figure}[H] 
\centering
\includegraphics[width=8.5cm,height=4.0 cm]{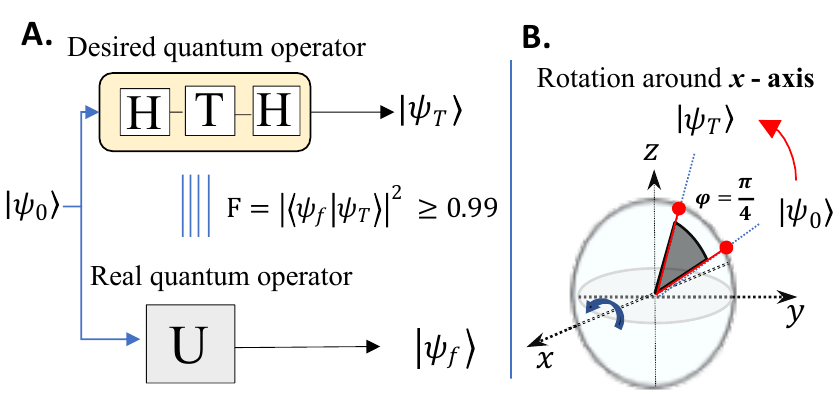}
\caption{(A) Gate design for single-qubit operator T$_{\text{x}}$ represented by a sequence of quantum operators $\text{HTH}$. The goal is to calibrate the real quantum operator $\text{U}$ to T$_{\text{x}}$ by increasing the fidelity F between the target quantum state $\ket{\psi{_T}}$ and the real quantum state $\ket{\psi{_f}}$. (B) The T$_{\text{x}}$ operator cause a rotation for the input state$\ket{\psi_0}$ around x-axis by $\frac{\pi}{4}$ as shown on the Bloch sphere.} \label{Tx_gate}
\end{figure}

\begin{figure}[H]
\centering
\includegraphics[width=9.3cm,height=7.4 cm]{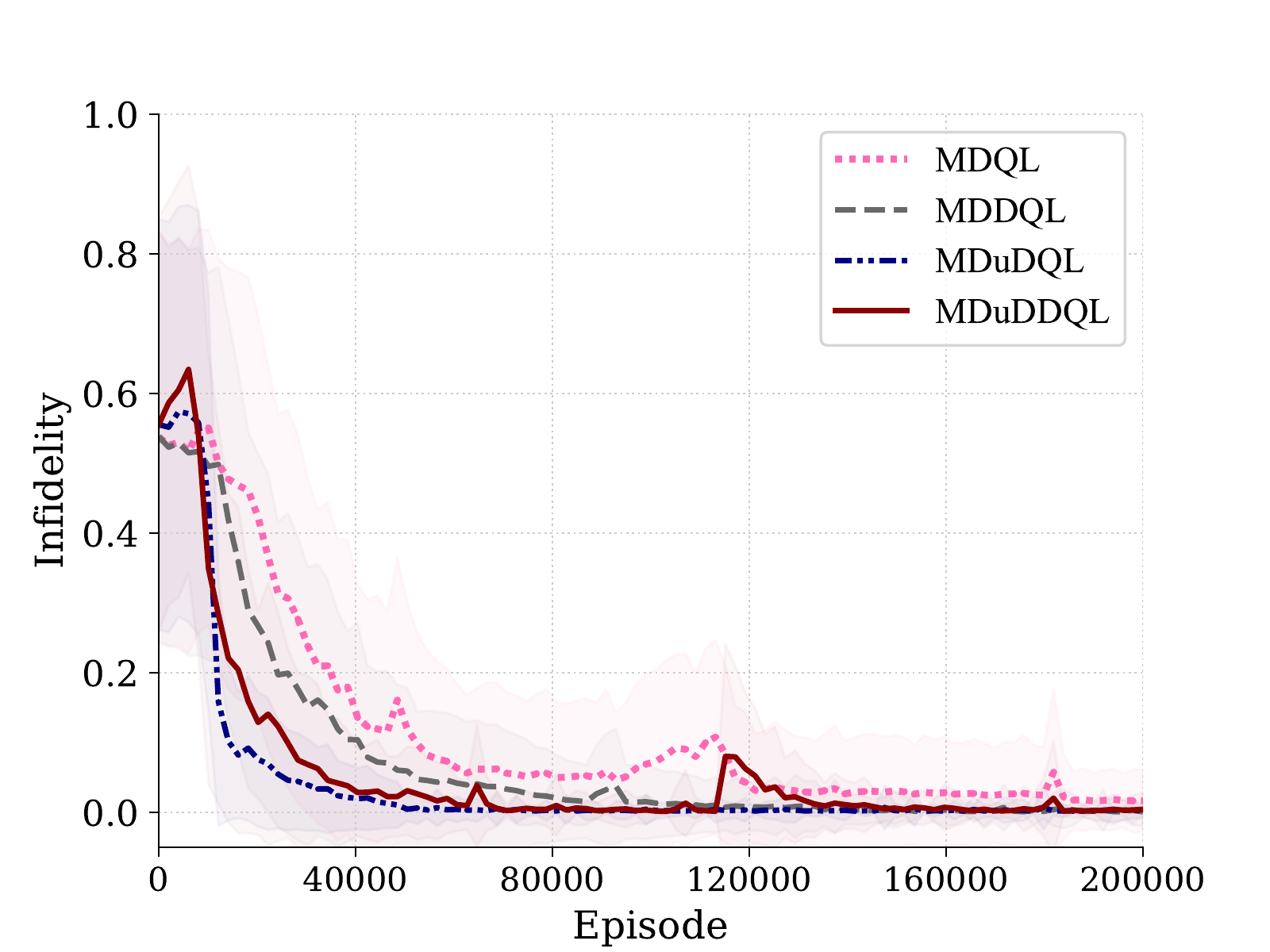}
\caption{The infidelity of the composed T$_{\text{x}}$ gate after calibration by using different RL algorithms. The solid lines represent the achieved average fidelity of 2000 samples for calibrating the quantum system to T$_{\text{x}}$ operator and the highlighted area represents the standard deviation.} \label{Train_Tx}
\end{figure}

\begin{figure}[H]
\includegraphics[width=8.9 cm,height=6.2cm]{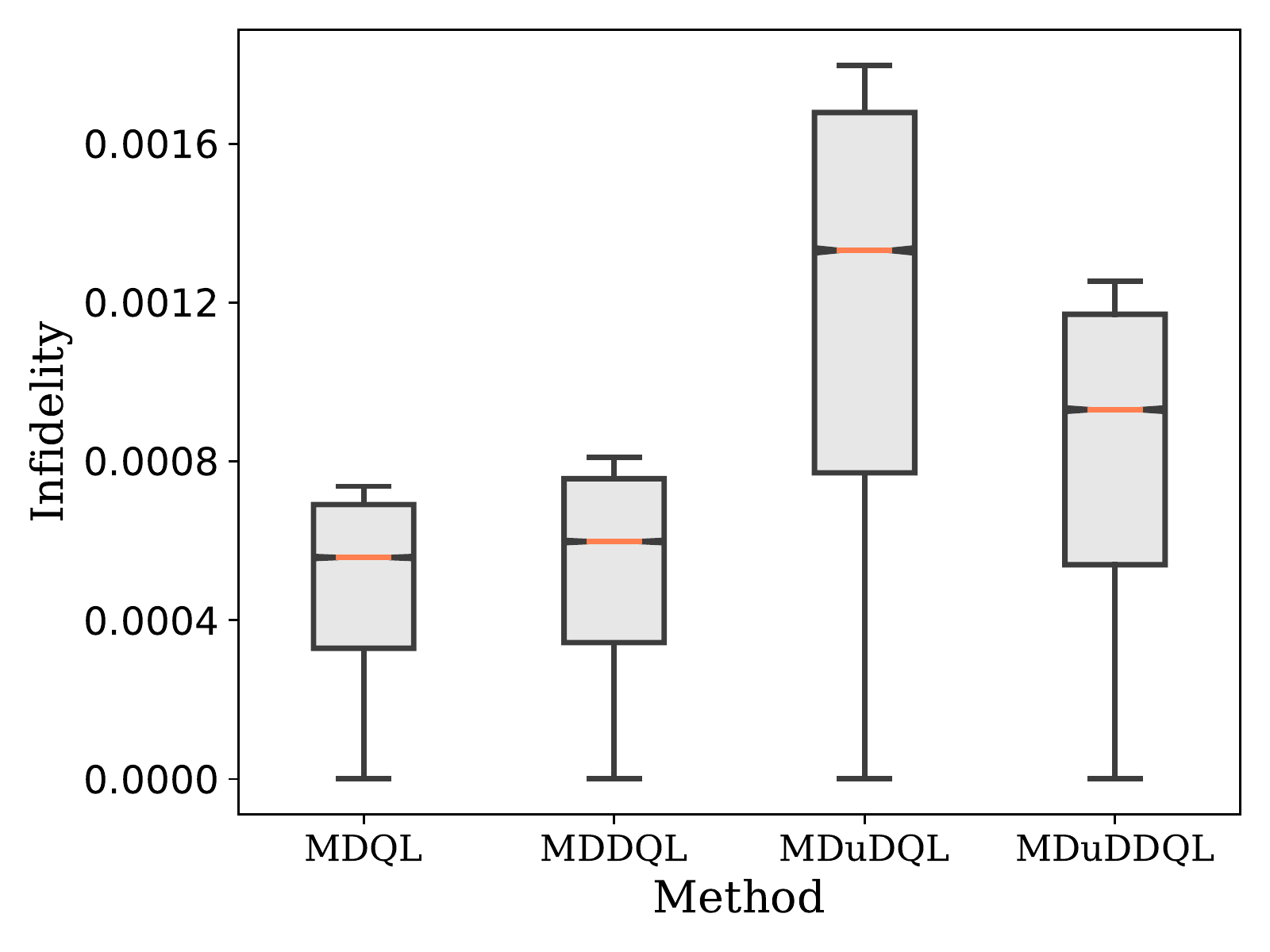}
\caption{Box plots showing testing infidelity of the calibrated composed single-qubit gate to T$_{\text{x}}$ operator by MDQL, MDDQL, MDuDQL and MDuDDQL for 50000 samples.} \label{Testing_Tx}
\end{figure}

Rotating a qubit around the x-axis by 45 degrees requires a sequence of three gates chosen from the \{H,T, S and CNOT\} gate-set (also known as the Clifford+T), as shown in Figure \ref{Tx_gate}. The goal is calibrating the applied gate U to the same operation as the T$_{\text{x}}$ gate. The model of the physical system of the applied gate is unknown. As seen in Figure \ref{Tx_gate}, known quantum states are utilised to test the performance of the applied gate for performing the desired operation. This process is also used to reward the DQL agent. 
For the training purpose, as explained in Appendix, we have used 100 quantum states for calibrating the gate. The training progress of the proposed algorithms is shown in Figure \ref{Train_Tx}. The DQL algorithms within the proposed framework succeed in calibrating the quantum gate and converging to good control policy.
To make sure the calibrated gate is unbiased to the training set, the calibrated gates are tested with 50000 samples, described in Appendix. The distribution of achieved infidelity for the testing set is presented via an interactive box plot \cite{Box_Plot1} shown in Figure \ref{Testing_Tx}. The horizontal lines within each box in the box plot graphing protocol stand in for the median, the upper and lower bounds of the interquartile range, and the whiskers, which indicate 1.5 times the interquartile range. In general, the infidelity of the worst case of the calibrated composed single-qubit gate to T$_{\text{x}}$ operator by the four algorithms is less than $10^{-2}$. This demonstrates the success of the model-free RL agent for the task.

\subsubsection{T$_{\text{y}}$ gate design}
\begin{figure}[H]
\centering
\includegraphics[width=8.5cm,height=4cm]{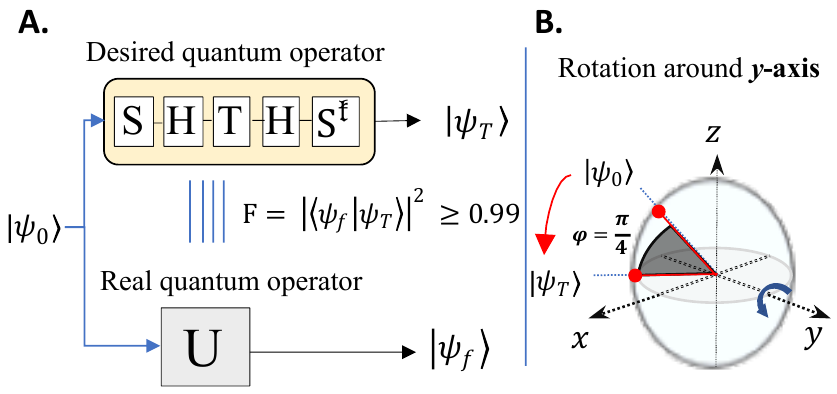}
\caption{(A) Gate design for single qubit operator T$_{\text{y}}$ represented by a sequence of quantum operators $\text{SHTHS}^{t}$. The goal is to calibrate the real quantum operator $\text{U}$ to T$_{\text{y}}$ by increasing the fidelity F between the target quantum state $\ket{\psi{_T}}$ and the real quantum state $\ket{\psi{_f}}$. (B) The T$_{\text{y}}$ operator causes a rotation for the input state$\ket{\psi_0}$ around y-axis by $\frac{\pi}{4}$ as shown on the Bloch sphere.} \label{Fig:Ty}
\end{figure}

Rotating the qubit around the y-axis by 45 degrees can be accomplished by a series of gates shown in Figure \ref{Fig:Ty}. The proposed algorithms are used to calibrate the gate U to do the same as the gates in series as explained in the following figure.

\begin{figure}[H]
\centering
\includegraphics[width=9.3cm,height=7.6 cm]{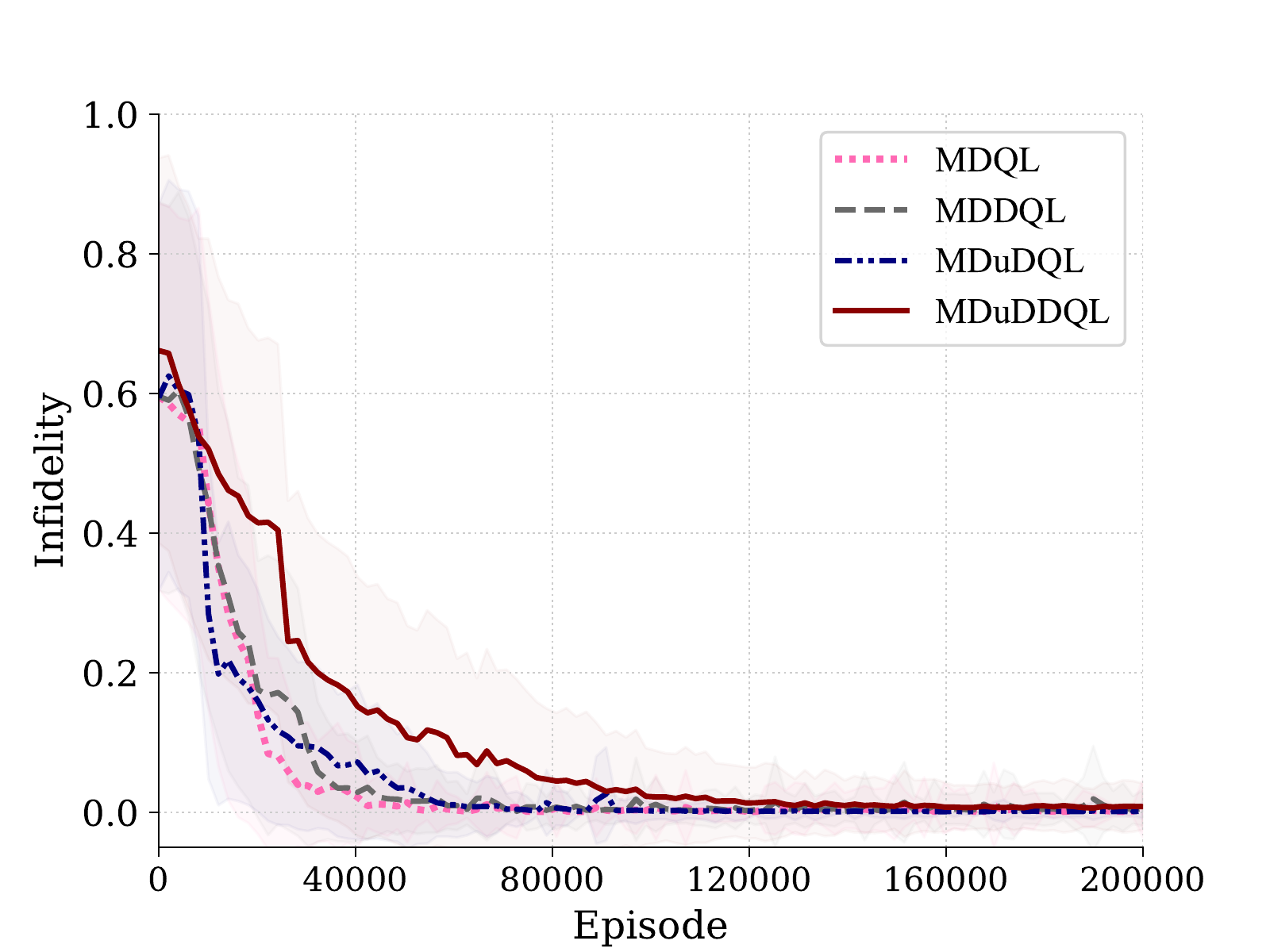}
\caption{The infidelity of the composed T$_{\text{y}}$ gate after calibration by using different RL algorithms. The solid lines represent the average achieved fidelity of 2000 samples of calibrating the quantum system to  T$_{\text{y}}$ gate. The highlighted area represents the standard deviation.}
\label{Fig:Ty_Training}
\end{figure}

The training progress for calibrating the U gate to T$_{\text{y}}$ operation is shown in Figure \ref{Fig:Ty_Training}. The DQL agent with the proposed approach succeeds in calibrating the U gate to T$_{\text{y}}$ gate.

\begin{figure}[H]
\includegraphics[width=8.9 cm,height=6.2cm]{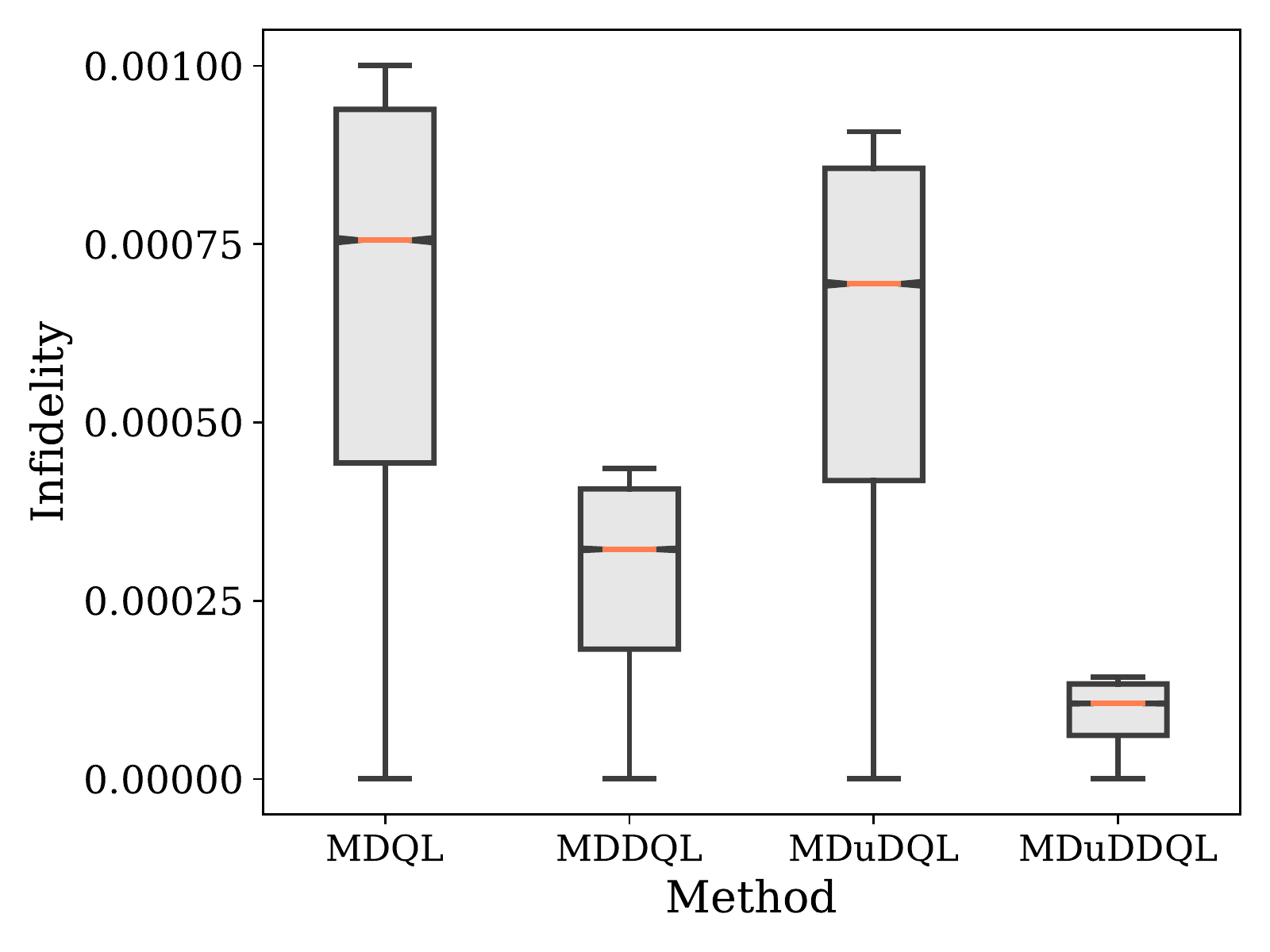}
\caption{Box plots showing testing infidelity of the calibrated composed single-qubit gate to T$_{\text{y}}$ operator by MDQL, MDDQL, MDuDQL and MDuDDQL for 50000 samples.} \label{Fig:Testing_Ty}
\end{figure}

As with the other gate calibration problems, the calibrated gate is tested for 50000 new samples, described in Appendix. The achieved infidelity for the testing set is shown in Figure \ref{Fig:Testing_Ty}. In general, the infidelity of the worst case of the calibrated composed single-qubit gate to T$_{\text{y}}$ operator by the four algorithms is less than $10^{-3}$. Based on the testing results, the agent of the proposed RL framework calibrates the single quantum gate to the desired operation successfully without any access to the dynamics of the quantum system. In a realistic quantum computer, the environmental decoherence of the qubits limits running a large-scale quantum algorithm. The suggested algorithm could benefit the quantum computer by reducing the size of the quantum circuit. 

\subsection{Quantum gate calibration within quantum circuit}

As explained in Section \ref{Calibration_gate1}, the goal for quantum gate calibration within a quantum circuit is to approximate the quantum gate to improve the worst fidelity $\min(\vec{F})$ of the training set. Two scenarios have been considered for quantum gate calibration. The first scenario for a single qubit system is Hadamard quantum gate calibration within a bit flip quantum circuit. The second scenario is CNOT quantum gate calibration within a Bell-state quantum circuit.

\subsubsection{Single qubit flip circuit} \label{Bit_flip}

\begin{figure}[H]
\centering
\includegraphics[width=6.5 cm,height=1 cm]{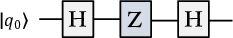}
\caption{Quantum circuit for bit flip of single qubit system.} \label{Fig:Figure 5}
\end{figure}

The quantum circuit shown in Figure \ref{Fig:Figure 5}, is used to flip the bit value of the input qubit. For example, if the initial state of $q_0$ is $\ket{0}$, the output state will equal $\ket{1}$. Z is a Pauli gate that causes a $180^o$ rotation of the qubit around the z-axis, while the gates with symbol H are the Hadamard gates. In this circuit, the goal for the DQL agent is to calibrate Hadamard gates.

\begin{figure}[H]
\centering
\includegraphics[width=9.3cm,height=7.4 cm]{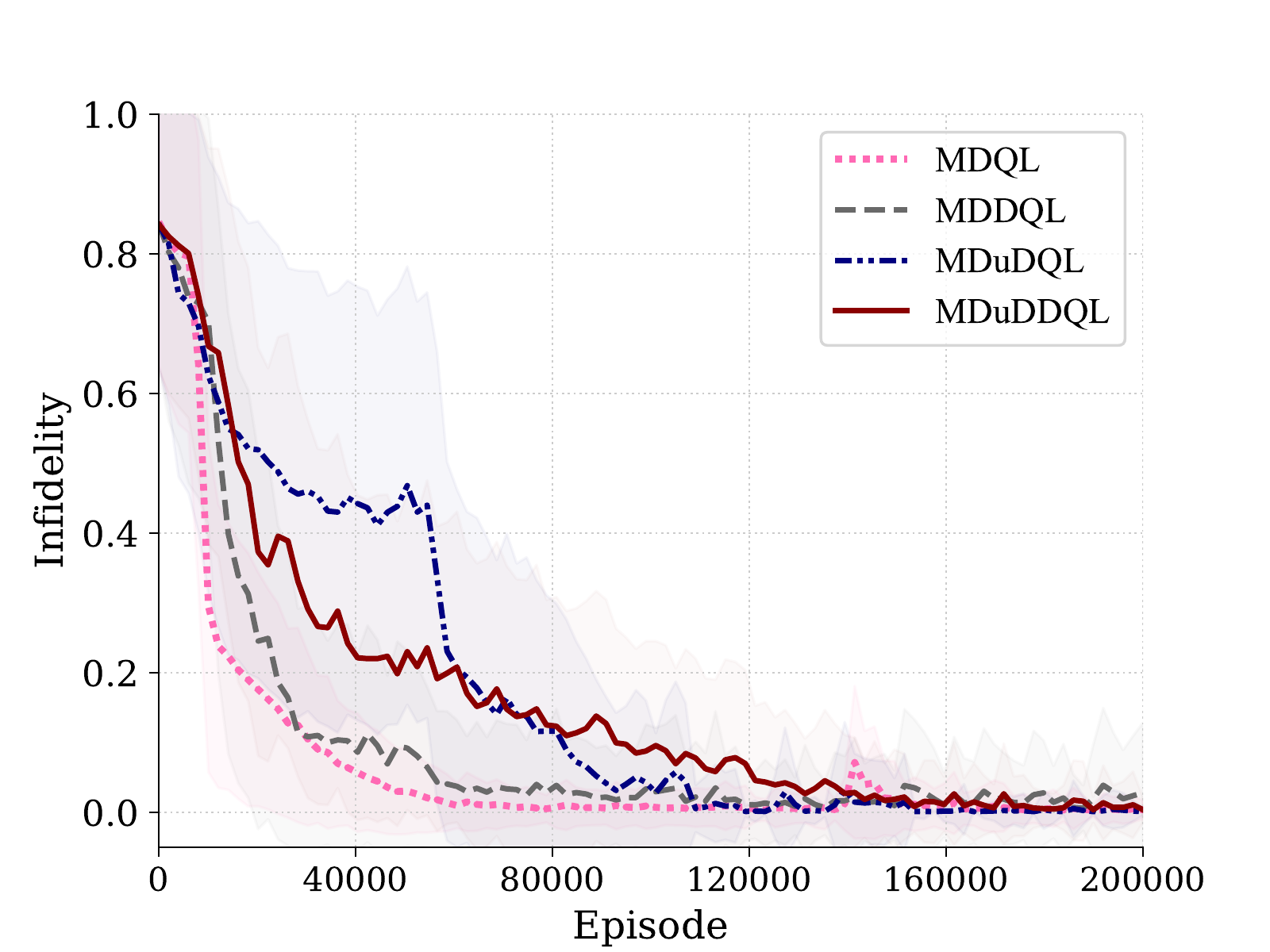}
\centering
\caption{The training accuracy of calibrating the gates to Hadamard gate within qubit flip circuit. The solid lines represent the average achieved fidelity of 2000 samples of calibrating the quantum system to Hadamard gate within qubit flip circuit. The highlighted area represents the standard deviation.} \label{Fig:Figure 6}
\end{figure}

Figure \ref{Fig:Figure 6} shows the average results of the worst achieved infidelity $(1 - \min(\vec{F}))$ of the outputs of the quantum circuit with the calibrated gates for the training quantum states. The size of the training set is 100 quantum states, the preparation of the training states is explained in Appendix. The control parameters like the action space, evolution time and number of control pulses are the same as in Section \ref{section 1}. As shown in Figure \ref{Fig:Figure 6}, MDQL, MDDQL, MDuDQL and MDuDDQL have succeed to calibrate the quantum gates in Figure \ref{Fig:Figure 5} to get the worst infidelity of the training set to less than $10^{-2}$.

To test whether the calibrated gates within the quantum circuit are similar to the desired gate and are not biased to the training set, 50000 new quantum states have been used to test approximated gates within the quantum circuit. The preparation of the testing states are shown in Appendix. Figure \ref{Fig:Figure 7} shows the infidelity results for the testing set for the best approximated gates by each method. It is worth noting that the upper dash represents the worst case, the lower dash represents the best case while the middle dash representing the median. In general, the infidelity of the worst case of the approximated Hadamard gates by the four algorithms is around $10^{-3}$. This is indicative that the model-free RL agent of MDQL, MDDQL, MDuDQL and MDuDDQL are successful to approximate the quantum gates of the single qubit system within the quantum circuit to the desired gates. 

\begin{figure}[H]
\includegraphics[width=8.9 cm,height=6.0cm]{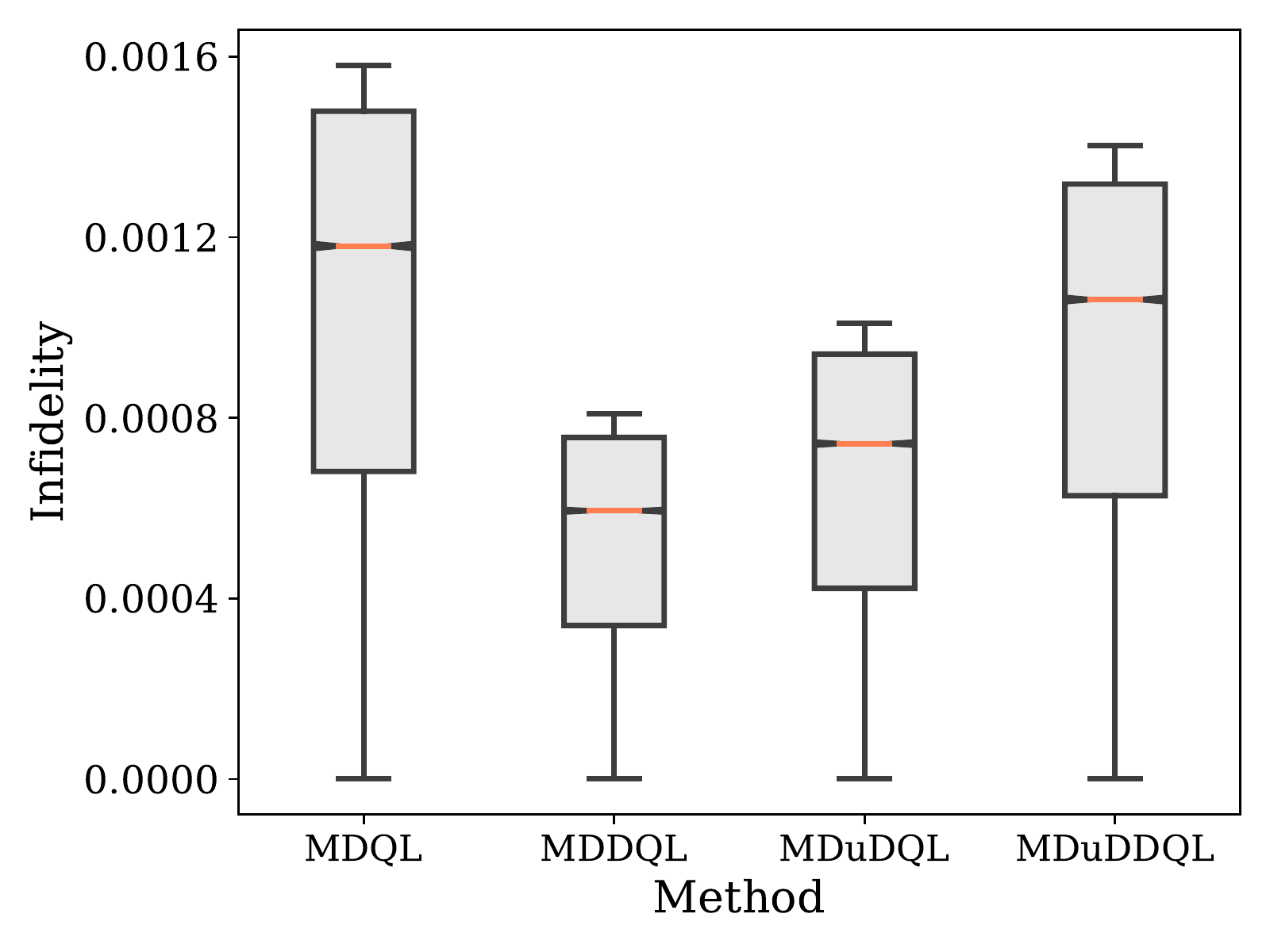}
\centering
\caption{Box plots showing testing infidelity of the single qubit filp circuit with the calibrated Hadamard gate by MDQL, MDDQL, MDuDQL and MDuDDQL for 50000 samples.} \label{Fig:Figure 7}
\end{figure}

\subsubsection{2-qubit  Bell state}\label{Bell_state}

\begin{figure}[H]
\centering
\includegraphics[width=4cm,height=3cm]{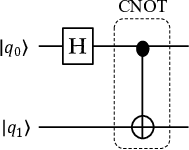}
\caption{The Bell state quantum circuit, that can be constructed by utilising a two-qubit circuit with a Hadamard gate on first qubit $\ket{q_0}$ and a CNOT gate on two qubits.}  \label{Fig:Figure 8}
\end{figure}

The quantum circuit shown in Figure \ref{Fig:Figure 8} is called Bell state circuit. It contains two gates, the Hadamard gate and the CNOT gate. This quantum circuit is used to generate correlated entangled quantum states. The first qubit $q_0$ is called the control qubit, while the second qubit $q_1$ is called the target qubit. The goal for the RL agent of the model-free quantum control method is to calibrate the CNOT gate within the Bell-state circuit. The control parameters of the RL algorithm are the same used in Section \ref{section 2}.

\begin{figure}[H]
\includegraphics[width=9.2 cm,height=7.6 cm]{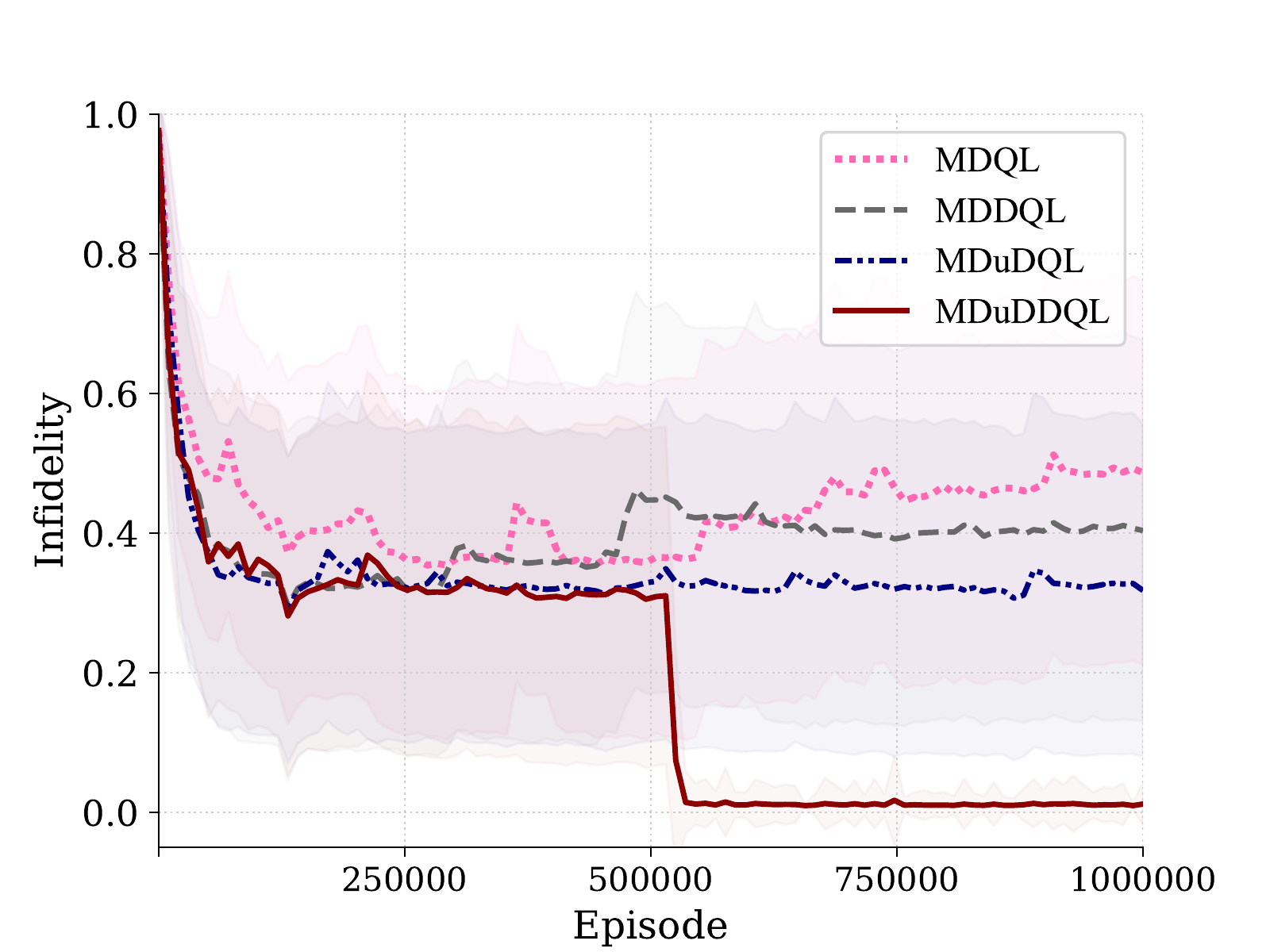}
\caption{The training accuracy of calibrating the gate to CNOT gate within Bell state quantum circuit. The solid lines represent the average achieved fidelity of 10000 samples of calibrating the quantum system to CNOT gate within Bell state quantum circuit. The highlighted area represents the standard deviation.} \label{Fig:Figure 9}
\end{figure}

Figure \ref{Fig:Figure 9} shows the average results of the worst achieved infidelity $(1 - \min(\vec{F}))$ of the outputs of the Bell state circuit with the calibrated CNOT gate for the training set. The training progress using 50 training states is shown in Figure \ref{Fig:Figure 9}. The details of choosing the training states are explained in Appendix. As shown in Figure \ref{Fig:Figure 9}, only MDuDDQL among the four algorithms has succeeded to converge to better control policy with lower infidelity.

As in the single qubit flip circuit, the training progress is not enough to tell if the calibrated gate is approximated to the CNOT gate or not. The testing states are supposed to be different from the training states. The testing results of 50000 quantum states are shown in Figure \ref{Fig:Figure 10}. The preparation of the testing states is explained in Appendix.   

\begin{figure}[H]
\centering
\includegraphics[width=8.9 cm,height=7.5cm]{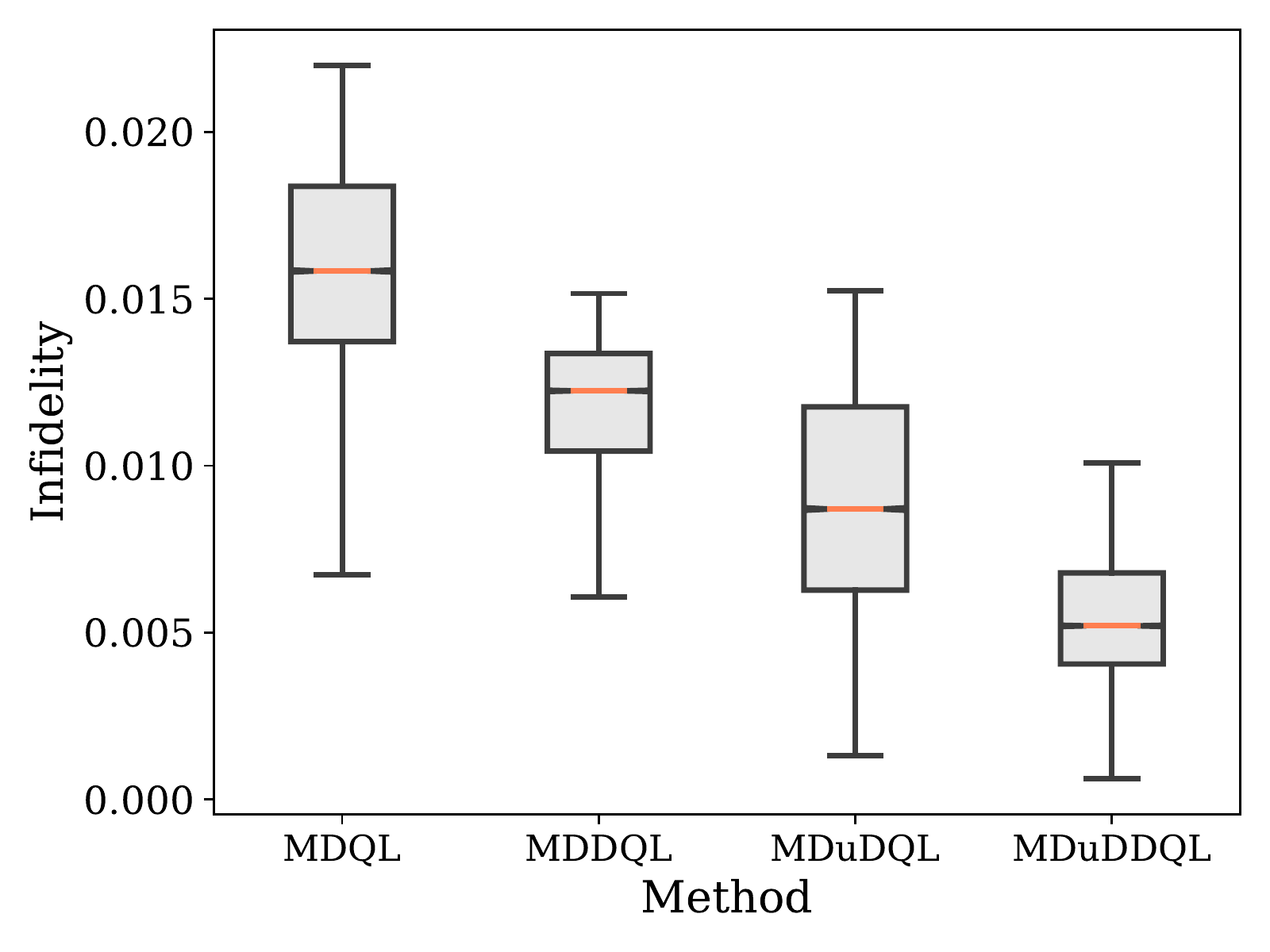}

\caption{Box plots showing testing infidelity of the Bell state circuit with the calibrated CNOT gate by MDQL, MDDQL, MDuDQL and MDuDDQL for 50000 samples.} \label{Fig:Figure 10}
\end{figure}

As shown in Figure \ref{Fig:Figure 10}, the worst achieved infidelity for MDuDDQL is less than $10^{-2}$, while the worst infidelities for MDQL, MDDQL and MDuDDQL are greater than $10^{-2}$. Also for the best case MDuDDQL has achieved the best results among the four algorithms. Based on the testing results the model-free frame works better with dueling double DQL algorithm for calibrating CNOT gate within a quantum circuit. This is due to combining the dueling networks and double estimation function that helps for reducing the overestimation and improving the exploration of the RL agent.

\section{Conclusion}\label{Conclusion}
    We have proposed a training framework for DQL algorithms that has been implemented with Dueling Double DQL (MDuDDQL), Double DQL (MDuDQL), Double DQL (DDQL) and MDQL to achieve model-free quantum gate design and quantum gate calibration. MDQL, MDDQL, MDuDQL and MDuDDQL succeeded in designing and calibrating the single qubit gates without any knowledge or access to the dynamics of the quantum system. MDDQL and MDuDDQL have shown better performance for quantum gate design and calibration for two qubit gates. The $n$-step reward function makes the state transitions of each episode unified by giving the same reward to all state transitions. This gives each state transition the ability to influence the RL agent when updating the prediction network and to take it towards better prediction policy. The modified experience replay memory keeps reminding the RL agent with the best discovered state transition experience to avoid any catastrophic drop in the performance. Quantum gate calibration could help reduce the requirements for a quantum algorithm and reduce the effort for error correction. The proposed framework seems promising for laboratory experiments involving quantum control, especially when the model of the quantum system is unknown or hard to find. This procedure may allow the DQL agent to work effectively even with large quantum systems without having to construct the Hamiltonian equation. In future work, we will compare the performance of on-policy reinforcement learning algorithms like Proximal Policy Optimization (PPO), and Deep Deterministic Policy Gradient (DDPG) with the proposed training framework for quantum gate design and calibration.
    

\setcounter{secnumdepth}{0}
\section{Appendix}\label{Appendix}

\subsection{A) Training and testing quantum states}\label{Appendix_B}

According to quantum mechanics, the state $\ket{\psi}$ of a single qubit in superposition can be represented as follows: 
\begin{equation}\label{1qubit_state}
\ket{\psi} = \alpha \ket{0} + \beta \ket{1}.
\end{equation}
The coefficients $\alpha \text{ and } \beta$ are the probability amplitudes of states $\ket{0} \text{and} \ket{1}$, respectively. $ \alpha \text{ and } \beta $ are complex numbers such that the state vector has length of one as follows:

\begin{equation}\label{Norm_1qubit_state}
\abs{\alpha}^2+\abs{\beta}^2 = 1
\end{equation}


For the single qubit systems in Sections \ref{Bit_flip} and \ref{U_single_gates} we use a quantum circuit to generate the training states. Each quantum circuit contains two Hadamard gates H and a phase gate $\varphi(\theta)$ in the sequence of [H $ \rightarrow \varphi(\theta) \rightarrow $ H]. This sequence of gates allows rotation of the qubit around the x-axis. The ground state $\ket{0}$ is considered as the initial input. The output state from the quantum circuit in the previous iteration is used as input state for the next iteration. The value of $\theta$ is equal to $0.16738\pi$ to avoid repeating any output state.

To generate a set of testing states different from the training states on the Bloch sphere, the following Hamiltonian is applied on a closed system,

\begin{equation}\label{Testing_states}
    H=\sigma_{z}+u\sigma_{x}.
\end{equation}

The quantum state in (\ref{1qubit_state}) evolves according to the Schr\"{o}dinger equation,
\begin{equation}\label{Schroedinger_states}
    \ket{\psi_{out}}= e^{(-iH(u)dt)}\ket{\psi_{in}}.
\end{equation}
The state $\ket{\psi_{out}}$ is the output quantum state for the evolution process of applying the control pulse $u$ for period  of time $dt$ with quantum state $\ket{\psi_{in}}$. The output states from the evolution process in (\ref{Schroedinger_states}) are the testing states. The testing states are used in Sections \ref{Bit_flip} and \ref{U_single_gates}. The value of the control pulse has been restricted to $u\in[4,-4]$ and the time of control pulse $dt = 0.05$. The quantum states have been generated by applying the evolution process iteratively as we have done for generating the training states. To discover a vast number of states on the Bloch sphere and to increase the randomness, the value of the control pulse has been chosen randomly. For the two-qubit system in Section \ref{Bell_state}, the training set has been chosen randomly from the testing set of the single-qubit system.

\end{sloppypar}

\end{document}